\documentclass[a4paper,11pt]{article}

\usepackage{amssymb,latexsym,amsmath,amsthm,amsfonts,graphics}
\def\prf{\vskip -0.1cm\noindent{\bf Proof\quad }}
\def\prfend{\hfill{$\Box$}\vskip 0.2cm}

\def\D{\{0,1\}}

\def\R{\mathbb{R}}

\newtheorem{lem}{Lemma}

\newtheorem{thm}{Theorem}

\theoremstyle{definition}
\newtheorem{dfn}{Definition}

\theoremstyle{remark}

\numberwithin{equation}{section}

\begin{document}
\parindent=1em
%------------------------------------ title ---------------------
\title{A characterization of eventually periodicity}
\author{Teturo Kamae
\and Dong Han Kim}

\date{}

\maketitle
%----------------------------------- abstract ------------------
\begin{abstract}
\noindent 
In this article, we show that the Kamae-Xue complexity function for an infinite sequence classifies eventual periodicity completely.
We prove that an infinite binary word $x_1x_2 \cdots $ is eventually periodic if and only if 
$\Sigma(x_1x_2\cdots x_n)/n^3$ has a positive limit,
where $\Sigma(x_1x_2\cdots x_n)$ is the sum of the 
squares of all the numbers of appearance of finite words in $x_1 x_2 \cdots x_n$, which was introduced by Kamae-Xue as a criterion of randomness in the sense that 
$x_1x_2\cdots x_n$ is more random if $\Sigma(x_1x_2\cdots x_n)$ is smaller. 
In fact, it is known that the lower limit of $\Sigma(x_1x_2\cdots x_n) /n^2 $ is at least 3/2  for any sequence $x_1x_2 \cdots$, while the limit exists as 3/2 almost surely for the $(1/2,1/2)$ product measure.
For the other extreme, the upper limit of $\Sigma(x_1x_2\cdots x_n)/n^3$ is bounded by 1/3.
There are sequences which are not eventually periodic but  
the lower limit of $\Sigma(x_1x_2\cdots x_n)/n^3$ is positive, while the limit does not exist. 
\end{abstract}

\section{Introduction}
In \cite{KX}, a criterion of randomness for binary words 
is introduced. As stated in Definition~\ref{def1} and \ref{def3}, 
let
$$\Sigma(x_1x_2\cdots x_n)=\sum_{\xi\in\cup_{k=1}^\infty\D^k}|x_1x_2\cdots 
x_n|_\xi^2,$$
where 
$$
|x_1x_2\cdots x_n|_\xi:=\#\{i : 1\le i\le n-k+1,~x_ix_{i+1}
\cdots x_{i+k-1}=\xi\}$$
is the number of appearance of a finite word $\xi$ in 
$x_1 x_2 \cdots x_n$. 
Since the function $f(x)=x^2$ is convex, the value 
$\sum_{\xi\in\D^k}|x_1x_2\cdots x_n|_\xi^2$ for any 
$k=1,2,\cdots$ is smaller if the values 
$|x_1x_2\cdots x_n|_\xi$ for $\xi\in\D^k$ are less deviated as a whole from the mean value $(n-k+1)/2^k$, that is, 
the sequence $x_1x_2\cdots x_n$ is more random. 
In fact, it is proved in \cite{KX} that 
$$
\liminf_{n\to\infty} \frac {\Sigma(x_1x_2\cdots x_n)}{n^2} 
\ge \frac 32
$$
holds for any $x_1x_2\cdots\in\{0,1\}^\infty$, while 
$$
\lim_{n\to\infty} \frac {\Sigma(X_1X_2\cdots X_n)}{n^2} = \frac 32
$$
holds with probability 1 if $X_1X_2\cdots X_n$ is the 
i.i.d. process with $P(X_i=0)=P(X_i=1)=1/2$.

In this article, we study the opposite case that 
$\Sigma(x_1x_2\cdots x_n)$ increase in the order of 
$n^3$ and prove that 
$x_1x_2\cdots \in\{0,1\}^\infty$ is eventually periodic if and only if 
$$
\lim_{n\to\infty} \frac  {\Sigma(x_1x_2\cdots x_n)}{n^3}
\mbox{ exists and }>0.
$$
It is easy to see that if $x=x_1x_2\cdots\in\D^\infty$ 
containes a few 1, or exactly speaking, if 
$x=0^{k_1}10^{k_2}1\cdots$ with 
$\liminf_{n\to\infty}k_{n+1}/k_n>1$, then we have 
$$
\liminf_{n\to\infty} \frac {\Sigma(x_1x_2\cdots x_n)}{n^3}>0.
$$
Since this $x_1x_2\cdots$ is not eventually periodic, 
it follows from our result that  
$\lim_{n\to\infty} \Sigma(x_1x_2\cdots x_n)/n^3$ 
does not exist. 

There are many characterizations of eventually 
periodicity. 
Most famous one might be the result due to Hedlund
and Morse concerning the complexity. That is, 
$x_1x_2\cdots$ is eventually periodic 
if and only if for some $k\ge 1$ the number of words of 
size $k$ appearing in $x_1x_2\cdots$ is smaller than 
$k+1$ (\cite{MoHe}). Another characterization concerning 
the return time is obtained in \cite{Ki}. 
Here, we add one more characterization which concerns 
both the complexity and the return time. 

\section{Definitions and Lemmas}

\begin{dfn}\label{def1}
For $x_1x_2\cdots x_n\in\{0,1\}^n$, $\xi\in\D^k$ with 
$1\le k\le n$ and $i=0,1,\cdots,n-k$, we denote 
$$\xi\prec_i x_1x_2\cdots x_n
~\mbox{ if }~\xi=x_{i+1}x_{i+2}\cdots x_{i+k}$$ 
and 
$$
\xi\prec x_1x_2\cdots x_n
~\mbox{ if }~\xi \prec_i x_1x_2\cdots x_n
\mbox{ for some }i=0,1,\cdots,n-k.
$$ 
We call $\xi$ a {\it factor} or {\it suffix} of 
$x_1x_2\cdots x_n$, 
respectively, if $\xi\prec x_1x_2\cdots x_n$ or 
$\xi\prec_{n-k} x_1x_2\cdots x_n$. 
We also denote 
\begin{equation*}
|x_1x_2\cdots x_n|_\xi=\#\{i : 0\le i\le n-k, \ \xi\prec_i x_1x_2\cdots x_n\} 
\end{equation*}
and $|x_1x_2\cdots x_n|=n$. 
\end{dfn}

\begin{dfn}
For $\eta= a_1 \cdots a_k\in\{0,1\}^k$ and 
$\ell=1,2,\cdots$, we denote 
$$\eta^\ell=\underbrace{a_1\cdots a_k}_1
\underbrace{a_1\cdots a_k}_2
\cdots\underbrace{a_1\cdots a_k}_\ell.$$
In the same way, we define $\eta^\infty\in\D^\infty$. 
We call $\eta$ {\it prime} if there is no $\xi$ 
such that $\eta=\xi^\ell$ for some $\ell\ge 2$.   
\end{dfn}

\begin{dfn}[\cite{KX}]\label{def3}
Define $\Sigma^n :\D^n \to\R$ by 
$$
\Sigma^n (x_1x_2\cdots x_n)=\sum_{\xi\in\D^+} |x_1x_2\cdots x_n|_\xi^2,
$$ 
where $\D^+=\bigcup_{k=1}^\infty\D^k$. 
We write $\Sigma^n = \Sigma$ as a function from $\D^+$ to $\R$.
\end{dfn}

\begin{dfn}
For $x_1x_2\cdots x_n\in\{0,1\}^n$, define
$$
\Lambda(x_1x_2\cdots x_n)=\max\{|\eta|^2(\ell+1)^3 : \eta^\ell\prec x_1x_2\cdots x_n\}$$
\end{dfn}

\begin{lem}\label{lem1}
For any $x_1x_2\cdots x_n\in\{0,1\}^n$, it holds that 
$$
\Sigma(x_1x_2\cdots x_n)\ge \frac {\Lambda(x_1x_2\cdots x_n)}{48}. 
$$
\end{lem}
\prf
Let $M=\Lambda(x_1x_2\cdots x_n)$. Then, there exist 
positive integers $k,\ell$ and $\eta\in\D^k$ with 
$\eta^l\prec x_1x_2\cdots x_n$ such that 
$k^2(\ell+1)^3=M$. 
Then, we have
$$\sum_{\xi;~\xi\prec\eta}
|\eta^\ell|^2_\xi\ge l\sum_{\xi;~\xi\prec\eta}|\eta^\ell|_\xi
\ge \frac{k^2\ell^2}{2}, 
$$
since $|\eta^\ell|_\xi\ge \ell$ if $\xi\prec\eta$ and 
$\sum_{\xi;~\xi\prec\eta}|\eta^\ell|_\xi\ge k^2\ell/2$. 
In the same way, for any $i=1,\cdots,\ell-1$, we have 
$$\sum_{\xi;~\xi\not\prec\eta^i~{\rm and}
~\xi\prec\eta^{i+1}}|\eta^\ell |_\xi^2 \ge \frac{k^2(\ell-i)^2}{2}. 
$$
Therefore, we have
\begin{equation*}\begin{split}
\Sigma(x_1x_2\cdots x_n) 
&\ge\sum_{i=0}^{\ell-1} \sum_{\xi;~\xi\not\prec\eta^i \text{ and } \xi\prec\eta^{i+1}}|\eta^\ell|_\xi^2 \\
&\ge\sum_{i=0}^{\ell-1} \frac {k^2(\ell-i)^2}{2}
\ge \frac{k^2 \ell^3}{6} \ge  \frac{k^2(\ell+1)^3}{48} = \frac {M}{48}.
\end{split}\end{equation*}
\prfend

\begin{lem}\label{lem2}
For any $x_1x_2\cdots\in\{0,1\}^\infty$, 
$$
\liminf_{n\to\infty} \frac{\Sigma(x_1x_2\cdots x_n)}{n^3} >0 
\text{ if and only if }
\liminf_{n\to\infty}\frac{\Lambda(x_1x_2\cdots x_n)}{n^3}>0,
$$
and
$$
\limsup_{n\to\infty}\frac {\Sigma(x_1x_2\cdots x_n)}{n^3}>0 
\text{ if and only if }
\limsup_{n\to\infty}\frac {\Lambda(x_1x_2\cdots x_n)}{n^3}>0.
$$
\end{lem}
\prf
By Lemma~\ref{lem1}, the ``if" parts are clear. Let us prove the 
``only if" parts. 
Let $M_n=\Lambda(x_1x_2\cdots x_n)$. 
Assume that there exists $i,~m,~k$ with 
$1\le i+1<i+k+1<i+m<i+k+m\le n$ such that 
$$x_{i+1}x_{i+2}\cdots x_{i+m}=x_{i+k+1}x_{i+k+2}\cdots x_{i+k+m}.$$
Let $k$ be the minimum with this property. 
Let $\eta=x_{i+1}x_{i+2}\cdots x_{i+k}$ and 
$\ell=\lfloor m/k\rfloor$. Then, 
$k^2(\ell+1)^3\le M_n$ holds since 
$\eta^\ell \prec x_{i+1}x_{i+2}\cdots x_{i+m}$. 
Hence, $k\ge (k(\ell+1))^3/M_n>m^3/M_n$. 
It follows that  $|x_1x_2\cdots x_n|_\xi\le n/(m^3/M_n)$ 
for any $\xi\in\D^m$. Therefore for any $ 1 \le m \le n$, 
\begin{equation}\label{2.1}\begin{split}
\sum_{\xi\in\D^m}|x_1x_2\cdots x_n|^2_\xi
&\le \frac{n}{m^3/M_n}\sum_{\xi\in\D^m}|x_1x_2\cdots x_n|_\xi \\
&\le \frac{n}{m^3/M_n} \cdot n = \frac{n^2M_n}{m^3}.
\end{split}\end{equation}

Assume that 
$$
\liminf_{n\to\infty} \frac {M_n}{n^3} =0.
$$
Then, there exists a subsequence $\{n'\}$ of 
$\{n\}=\{1,2,\cdots\}$ such that 
$M_{n'}/{n'}^3$ converges to 0 as $n'\to\infty$. 
Let $\psi(n)=(M_n/n^3)^{1/3}$. Then, $\psi(n')\to 0$ 
as $n'\to\infty$. For simplicity, we denote this subsequence $\{n'\}$ by $\{n\}$.  
By \eqref{2.1}, we have 
\begin{align*}
\frac {\Sigma(x_1x_2\cdots x_n)}{n^3} 
&= \frac 1{n^3} \sum_{m=1}^n\sum_{\xi\in\D^m}
|x_1x_2\cdots x_n|^2_\xi\\
&\le \frac 1{n^3} \sum_{1\le m\le\psi(n)n}(n-m)^2
+\frac 1{n^3} \sum_{m>\psi(n)n} \frac{n^2M_n}{m^3} \\
&\le \frac 1{n^3} \cdot \psi(n)n\cdot n^2
+ \frac {M_n}{n} \cdot \frac 1 {2(\psi(n)n-1)^{2}}\\
&=\psi(n)+  \psi(n)^3 \cdot \frac 1{2\psi(n)^{2}} \cdot \frac{1}{(1-(\psi(n)n)^{-1})^{2}}\\
&=\psi(n)+\frac{\psi(n) }{2(1-(M_n)^{-1/3})^{2}},
\end{align*}
which implies that 
$$
\liminf_{n\to\infty} \frac {\Sigma(x_1x_2\cdots x_n)}{n^3} =0.
$$
By the same argument, we can prove that 
$$
\limsup_{n\to\infty}  \frac {M_n}{n^3} =0
$$
implies that 
$$
\limsup_{n\to\infty} \frac{\Sigma(x_1x_2\cdots x_n)}{n^3} =0,
$$
which completes the proof.
\prfend

\begin{dfn}
For $\omega \in\D^n$, $\xi\in\D^k$ with $k\le n$ and 
$m=1,2,\cdots,n$, we denote 
$$
|\omega |_{\xi,m}=\#\{i;~n-m-k+1\le i\le n-k,~\xi\prec_i \omega \}.
$$
\end{dfn}

\begin{lem}
Let $\omega \in\D^{n}$ and $\eta \in\D^{m}$ 
with $n,m\ge 1$. 
Then, we have 
$$
\Sigma (\omega\eta)-\Sigma (\omega)=\sum_{\xi\in\D^+} 2|\omega\eta|_{\xi,m} |\omega|_\xi
+|\omega \eta|^2_{\xi,m}.
$$
\end{lem}
\prf
Clear from the fact that 
$|\omega\eta|_\xi=|\omega|_\xi+|\omega\eta|_{\xi,m}$. 
\prfend

\begin{lem}\label{lem4}
Let $\omega\in\D^n$ and  $\eta \in\{0,1\}^k$ 
satisfy that $|\omega \eta^\ell|_{\eta^\ell}=1$. 
Assume that $\eta$ is prime and $\omega_n\ne\eta_k$ (i.e. 
the last elements of $\omega$ ad $\eta$ are different).
Then, for $\ell=2,3,\cdots$, we have 
\begin{equation}\label{2.2}
0 \le  \Sigma (\omega \eta^{\ell+2})-2\Sigma (\omega \eta^{\ell+1})+\Sigma (\omega \eta^\ell)
- 2k^2\ell < 2k^4 + 3k.
\end{equation}
\end{lem}

\prf
Put $\sigma = \omega \eta^{\ell}$. Denote 
\begin{equation*}
\begin{split}
\Sigma (\sigma\eta^2)-2\Sigma (\sigma\eta)+\Sigma (\sigma) 
=&\sum_{\substack{\xi\in\D^+; \\ \xi\prec\eta^{\ell+2}, |\xi|\ge k}}
|\sigma\eta^2|^2_\xi-2|\sigma\eta|^2_\xi+|\sigma|^2_\xi\\
&+\sum_{\substack{\xi\in\D^+;\\ \xi\prec\eta^{\ell+2}, |\xi|<k}} |\sigma\eta^2|^2_\xi-2|\sigma\eta|^2_\xi+|\sigma|^2_\xi\\
& +\sum_{\substack{\xi\in\D^+;\\ \xi\not\prec\eta^{\ell+2}}} |\sigma\eta^2|^2_\xi-2|\sigma\eta|^2_\xi+|\sigma|^2_\xi\\
=&: S_1+S_2+S_3.
\end{split}
\end{equation*}

Let $\xi\prec\eta^{\ell+1}$ with $|\xi|\ge k$. Since 
$\eta$ is prime, if $\xi\prec_i\eta^{\ell+1}$, then 
$\xi\prec_j\eta^{\ell+1}$ holds if and only if 
$i\equiv j \pmod k$ and $j+|\xi|\le|\eta^{\ell+1}|$. 
Therefore, $|\sigma\eta|_\xi-|\sigma|_\xi=|\sigma\eta|_{\xi,k}=1$. Hence, 
$|\sigma\eta|^2_\xi-|\sigma|^2_\xi=2|\sigma|_\xi+1$. 
In the same way, $|\sigma\eta^2|^2_\xi-|\sigma\eta|^2_\xi=2|\sigma\eta|_\xi+1$. 
Thus, 
$$
|\sigma\eta^2|^2_\xi-2|\sigma\eta|^2_\xi+|\sigma|^2_\xi=2(|\sigma\eta|_\xi-|\sigma|_\xi)
=2.
$$
If $\xi\prec\eta^{\ell+2}$ but not $\xi \prec \eta^{\ell+1}$, 
then by the assumptions that $|\omega \eta^\ell|_{\eta^\ell}=1$, $\eta$ is prime and $\omega_n\ne\eta_k$, $|\sigma\eta^2|_\xi=1$ and 
$|\sigma\eta|_\xi=|\sigma|_\xi=0$ hold. Hence,   
$$
|\sigma\eta^2|^2_\xi-2|\sigma\eta|^2_\xi+|\sigma|^2_\xi=1. 
$$
Therefore, 
$$
S_1=2((\ell k+1)+\ell k+\cdots+(\ell k-k+2))+k^2=2k^2 \ell+3k,
$$
since the number of $\xi\prec\eta^{\ell+1}$ with 
$|\xi|\ge k$ is equal to the number of the pairs of 
positions $(i,j)\in\{1,2,\cdots,(\ell+1)k\}^2$ in $\eta^{\ell+1}$ with $1\le i\le k$ and $j-i\ge k$. Also, the number of 
$\xi$ with $\xi\not\prec\eta^{\ell+1}$ and $\xi\prec\eta^{\ell+2}$ is equal to the number of 
the pairs of positions $(i,j)\in\{1,2,\cdots,(\ell+2)k\}^2$ 
in $\eta^{\ell+2}$ with $1\le i\le k$ and  $(\ell+1)k+1\le j\le (\ell+2)k$. 

Let $\xi\prec\eta^{l+2}$ with $|\xi|<k$. Then, 
$1\le|\sigma|_{\xi,k}=|\sigma\eta|_{\xi,k}=|\sigma\eta^2|_{\xi,k}\le k$ 
and $|\sigma\eta^2|_\xi-|\sigma\eta|_\xi=|\sigma\eta|_\xi-|\sigma|_\xi=|\sigma|_{\xi,k}$. 
Hence,
\begin{equation*}
\begin{split}
|\sigma\eta^2|^2_\xi-2|\sigma\eta|^2_\xi+|\sigma|^2_\xi
&=(|\sigma\eta^2|^2_\xi-|\sigma\eta|^2_\xi)
-(|\sigma\eta|^2_\xi-|\sigma|^2_\xi)\\
&=(2|\sigma\eta|_\xi|\sigma\eta^2|_{\xi,k} 
+|\sigma\eta|_{\xi,k}^2)-(2|\sigma|_\xi|\sigma\eta|_{\xi,k}+|\sigma|_{\xi,k}^2) \\
&=2(|\sigma\eta|_\xi-|\sigma|_\xi)|\sigma|_{\xi,k}
=2|\sigma|^2_{\xi,k}. 
\end{split}
\end{equation*}
Therefore, $0\le S_2 < 2k^4$. 

If $\xi\prec \sigma$ with $\xi\not\prec\eta^{\ell+2}$,  
then it holds that $|\sigma|_\xi=|\sigma\eta|_\xi$ since 
$|\sigma\eta|_{\xi,k}=0$ by the assumptions that $|\omega \eta^\ell|_{\eta^\ell}=1$, $\eta$ is prime and $\omega_n\ne\eta_k$. If $\xi\not\prec \sigma$, $\xi\prec \sigma\eta$ and 
$\xi\not\prec\eta^{\ell+2}$, then we have 
$|\sigma|_\xi=0$ and $|\sigma\eta|_\xi=1$. Hence, 
$$
\sum_{\substack{\xi\in\D^+;\\ \xi\not\prec\eta^{\ell+2}}}
(|\sigma\eta|^2_\xi-|\sigma|^2_\xi)
=\#\{\xi\in\D^+;~\xi\not\prec\eta^{l+2},
~\xi\prec \sigma\eta,~\xi\not\prec \sigma\}=kn
$$
In the same way, we have 
$$
\sum_{\substack{\xi\in\D^+;\\ \xi\not\prec \eta^{\ell+2}}}
(|\sigma\eta^2|^2_\xi-|\sigma\eta|^2_\xi)=kn.
$$
Therefore, we have $S_3=0$. 

Thus, we have 
\begin{equation*}
0 \le \Sigma (\sigma\eta^2) - 2 \Sigma(\sigma\eta)+\Sigma (\sigma) - 2k^2\ell < 2k^4+3k.
\end{equation*}
\prfend

\begin{lem}\label{lem5}
Assume that 
$$
\limsup_{n\to\infty} \frac{\Sigma(x_1x_2\cdots x_n)}{n^3} >0.
$$
Then, there exists a prime $\eta\in\D^+$ and $0\le \ell_1\le \ell_2\le\cdots$
such that 
$$
\eta^{\ell_n}\prec x_1x_2\cdots x_n~\mbox{ and }~
\limsup_{n\to\infty} \frac{\ell_n}{n} >0.$$ 
\end{lem}

\prf
By Lemma~\ref{lem2}, we have 
$$
\limsup_{n\to\infty} \frac {\Lambda(x_1x_2\cdots x_n)}{n^3} >0.
$$
Hence, there exist $
\eta_n \in\D^+$ and $h_n$ 
for any sufficiently large $n$ with 
${\eta_n}^{h_n}\prec x_1x_2\cdots x_n$ such that 
$$\limsup_{n\to\infty} \frac{|\eta_n|^2h_n^3}{n^3} 
\ge\limsup_{n\to\infty} \frac {|\eta_n|^2(h_n+1)^3}{8n^3} >0.$$
Since $|\eta_n|^2h_n^3/n^3 \le 1/|\eta_n|$, 
$\liminf_{n\to\infty}|\eta_n|<\infty$. Therefore, 
there exist $\eta\in\D^+$ and 
$0 \le \ell_{1} \le \ell_{2} \le \cdots$ 
such that $\eta^{\ell_n}\prec x_1x_2\cdots x_n$ and 
$$ \limsup_{n\to\infty} \frac{\ell_n}{n}
= \frac{1}{|\eta|^{2/3}} \left( \limsup_{n\to\infty} \frac{|\eta|^2\ell_n^3}{n^3} \right)^{1/3}>0.$$ 
If $\eta$ is not prime 
and $\eta=\xi^p$ with a prime $\xi$, we may replace $\eta$ 
by $\xi$ and $\ell_n$ by $p \ell_n$. 
\prfend

\section{Main results}

\begin{thm}\label{thm1}
If $x = x_1x_2\cdots$ is eventually periodic with the 
minimal period $k$. Then, it holds that
$$
\lim_{n\to\infty} \frac{\Sigma (x_1x_2\cdots x_n)}{n^3} = \frac{1}{3k}. 
$$  
\end{thm}
\prf
Let $\eta\in\D^k$ be prime with $k \ge 1$. 
Let $x=\zeta\eta^\infty$ with 
$\zeta\in\D^+\cup\{\emptyset\}$, where $\emptyset$ is the 
empty word. Let $|\zeta|=h$. Then, 
for any $\xi\in\D^+$ with $|\xi|=\ell$, we have
$$0\le|\zeta\eta^n|_\xi-|\eta^n|_\xi\le h.$$
Hence, 
$$0\le|\zeta\eta^n|^2_\xi-|\eta^n|^2_\xi\le 
h(|\zeta\eta^n|_\xi+|\eta^n|_\xi).$$
Therefore, 
$$
0\le\sum_{\xi\in\D^\ell}|\zeta\eta^n|^2_\xi-
\sum_{\xi\in\D^\ell}|\eta^n|^2_\xi\le 
h((h+kn-\ell+1)+(kn-\ell+1))
$$
for any $\ell$ with $1\le \ell\le kn$, 
$$
0\le\sum_{\xi\in\D^\ell}|\zeta\eta^n|^2_\xi-
\sum_{\xi\in\D^\ell}|\eta^n|^2_\xi\le 
h(h+kn-\ell+1)
$$
for any $\ell$ with $kn+1\le \ell\le h+kn$,
and
$$\sum_{\xi\in\D^\ell}|\zeta\eta^n|^2_\xi-
\sum_{\xi\in\D^\ell}|\eta^n|^2_\xi=0$$ 
otherwise. 
Hence, 
$$
0\le\Sigma (\zeta\eta^n)-\Sigma(\eta^n)\le h(h+kn)(h+kn+1).
$$
Thus, 
$$
\lim_{n\to\infty} \frac{\Sigma (\zeta\eta^n)}{n^3} 
=\lim_{n\to\infty}  \frac{\Sigma (\eta^n)}{n^3} 
$$
holds in the sense that if the limit exists in one side, 
then the limit exists in the other side, and they coincides. 
Now, we prove that 
$$\lim_{n\to\infty}  \frac{\Sigma (\eta^n)}{n^3} = \frac{k^2}{3},$$
which will complete the proof. 

Assume that $|\xi|\ge k$ and $\xi\prec_i\eta^n$. 
Since $\eta$ is prime, $\xi\prec_j\eta^n$ holds 
if and only if $i\equiv j \pmod k$ and $0 \le j \le|\eta^n|-|\xi|$. 
Hence, for $\xi\prec\eta^n$ such that $|\xi|\ge k$, 
we have 
$$-1\le |\eta^n|_\xi-(n-|\xi|/k)\le 1.$$   
Therefore, it holds that 
\begin{equation*}\begin{split}
&\left|\sum_{\xi\in\D^+;~\xi\prec\eta^n,
~|\xi|\ge k}|\eta^n|^2_\xi~
-\sum_{\xi\in\D^+;~\xi\prec\eta^n,
~|\xi|\ge k} \left( n- \frac{|\xi |}{k} \right)^2\right|\\
&=\left|\sum_{\xi\in\D^+;~\xi\prec\eta^n,~|\xi|\ge k}
\left( |\eta^n|_\xi \, - \left(n - \frac{|\xi|}{k} \right) \right)\left( |\eta^n|_\xi \, + \left (n- \frac{|\xi|}{k} \right) \right) \right|\\
&\le \sum_{\xi\in\D^+;~\xi\prec\eta^n,
~|\xi|\ge k} \left(|\eta^n|_\xi \,+ \left(n- \frac{|\xi|}{k} \right)\right)\\
&\le \sum_{\xi\in\D^+;~\xi\prec\eta^n,
~|\xi|\ge k}(2|\eta^n|_\xi~+1)\le 2(kn)^2+k(kn)\le 3(kn)^2
\end{split}\end{equation*} 
On the other hand, if $\xi\prec\eta^n$ and $|\xi|<k$, 
then we have $1\le|\eta^n|_\xi\le kn$ and there are 
at most $k^2$ number of $\xi$ as this. 
Therefore, 
$$
0\le\Sigma(\eta^n)-\sum_{\xi\in\D^+;~\xi\prec\eta^n,
~|\xi|\ge k}|\eta^n|^2_\xi\le k^2(kn)^2.$$

Thus,
$$\lim_{n\to\infty} \frac {\Sigma (\eta^n)}{n^3}
=\lim_{n\to\infty} \frac 1{n^3} \sum_{\xi\in\D^+;~\xi\prec\eta^n,~|\xi|\ge k} \left( n- \frac{|\xi|}{k} \right)^2.$$
Here, $\xi$ as above corresdopnds to the 
pair $(i,j)$, where $i$ is the smallest $i$ such that 
$\xi\prec_{i-1}\eta^n$ and $|\xi|=j$. 
This correspondence gives an bijection between the set of $\xi$ 
as above and the set 
$$\{(i,j)\in\{1,2,\cdots,k\}\times\{k,k+1,\cdots,kn\};
~i+j-1\le kn\}. $$ 
Hence, we have
\begin{equation*}\begin{split}
\lim_{n\to\infty}\frac{\Sigma (\eta^n)}{n^3}
&=\lim_{n\to\infty} \frac 1{n^3}\sum_{\xi\in\D^+;
~\xi\prec\eta^n,~|\xi|\ge k} \left(n- \frac{|\xi|}{k} \right)^2\\
&=\lim_{n\to\infty} \frac 1{n^3}\sum_{(i,j)\in
\{1,2,\cdots,k\}\times\{k,k+1,\cdots,kn\},~i+j-1\le kn}
\left(n- \frac {j}{k} \right)^2\\
&=\lim_{n\to\infty} \frac{k}{n^3}
\sum_{j\in\{k,k+1,\cdots,kn\}} \left( n - \frac jk \right)^2\\
&=k^2\lim_{n\to\infty} \sum_{j\in\{1,2,\cdots,kn\}} \left(1-\frac{j}{kn} \right)^2 \frac 1{kn} \\
&=k^2\int_0^1(1-x)^2dx= \frac{k^2}{3}. 
\end{split}\end{equation*}
\prfend

\begin{thm}
It holds that $x_1x_2\cdots$ is eventually periodic 
if and only if $\lim_{n\to\infty} \Sigma(x_1x_2\cdots x_n)/n^3 $ exists and take a positive 
value.  
\end{thm}
\prf
The ``only if'' part is proved in Theorem~\ref{thm1}. Let us 
prove the ``if'' part. 

Suppose that 
$\lim_{n\to\infty} \Sigma(x_1x_2\cdots x_n)/n^3$
exists and takes a positive value, but $x_1x_2\cdots$ 
is not eventually periodic. 

By Lemma~\ref{lem5}, there exist $k\ge 1$,  
$\eta  = a_1 a_2 \cdots a_k \in \D^k$ and  
$0\le \ell_1\le l_2\le\cdots$ such that 
$$\eta^{\ell_n}\prec x_1x_2\cdots x_n~
\mbox{ and }~ A:=\limsup_{n\to\infty} \frac{\ell_n}{n} >0.$$ 
Here, we may also assume that $\eta$ is prime.

Take a subsequence $\{N\}$ of $\{1,2,\cdots\}$ and replace 
$\eta$ by $a_i \cdots a_k a_1\cdots a_{i-1}$ 
for some $i$ with $1\le i\le k$
if necessary, we may assume that $\eta^{\ell_N}$ is a 
suffix of $x_1x_2\cdots x_N$ and $x_{N-k\ell_N} \ne a_k$. 
Since $x_1x_2\cdots$ is not eventually periodic, we may 
also assume that $N-k\ell_N\to\infty$ as $N\to\infty$. 
Note that $kA\le 1$. 

Take $\delta>0$ with $1-kA<\delta<1$. Take $\epsilon$ 
with $0<\epsilon<1/2$ such that 
$(1-kA(1-\epsilon))(1+\epsilon)/(1-\epsilon)<\delta<1$. 
Take a sufficiently large $N$ such that $\delta \ell_N\ge 2$ 
and $\ell_N/N>A(1-\epsilon)$ together with other requirements 
specified later.  
 
We assume that $N-k\ell_N$ is sufficiently large. 
Denote $n=N-k\ell_N$ and $\omega=x_1x_2\cdots x_n$. 
Then, $x_n\ne a_k$. 
Since $n$ is sufficiently large, we may assume that 
$\ell_n/n<A(1+\epsilon)$. Hence, 
\begin{equation*}\begin{split}
\ell_n &<A(1+\epsilon)n=A(1+\epsilon)(N-k\ell_N)\\
&<A(1+\epsilon)(N-kAN(1-\epsilon))=(1+\epsilon) (1-kA(1-\epsilon))AN\\
&<(1+\epsilon)(1-kA(1-\epsilon)) \frac{\ell_N}{1-\epsilon} <\delta \ell_N. 
\end{split}\end{equation*}

Take integers $\ell$ and $\ell'$ as functions of $N$ such that
\begin{enumerate}
\renewcommand*\labelenumi{(\theenumi)}
\item\label{1} $\delta^{-1}\ell_n<\ell<\ell+2\ell'< \ell_N$ and
\item\label{2} $\ell/n$ and $\ell'/n$ are bounded away both from 0 and $\infty$.
\end{enumerate}
Since $x_n\ne a_k$ and $\eta$ is prime, $|x\eta^\ell|_{\eta^\ell}\ge 2$ is possible 
only if $|x\eta |_{\eta^\ell}\ge 1$, and hence, only if 
$|x|_{\eta^{\ell-1}}\ge 1$. This is impossible since 
$\ell_n<\delta \ell<\ell-1$ as $\delta<1$ and $N$ is sufficiently 
large. Thus, the assumptions in Lemma 4 are satisfied. 

Adding \eqref{2.2} for $m=k\ell,k(\ell+1),\cdots,k(\ell+\ell'-1)$, 
we have 
\begin{equation}\label{3.1}
\begin{split}
&\Sigma(\omega\eta^{\ell+\ell'+1})
-\Sigma(\omega\eta^{\ell+\ell'})
-\Sigma(\omega\eta^{\ell+1})
+\Sigma(\omega\eta^{\ell})\\
&=2k^2(\ell+(\ell+1)+\cdots+(\ell+\ell'-1))+\ell' R\\
&=k^2\ell'(2\ell+\ell'-1)+\ell' R
\end{split}
\end{equation}
for some $R$ with $0\le R < 2k^4+3k$.
 
We further add \eqref{3.1} for the pairs 
$(\ell,\ell'),(\ell+1,\ell'),\cdots,(\ell+\ell'-1,\ell')$ in place of $(\ell,\ell')$,  
we get 
\begin{equation}\label{3.2} \begin{split}
\Sigma(\omega\eta^{\ell+2\ell'}) -2\Sigma(\omega\eta^{\ell+\ell'}) +\Sigma(\omega\eta^{\ell}) 
&=\sum_{i=l}^{\ell+\ell'-1} \left(k^2\ell'(2i+\ell'-1)+\ell' R_i \right) \\
&=2k^2 \ell'^2 (\ell+\ell'-1)+\ell'^2 \bar R
\end{split}\end{equation}
with some $\bar R$, $0 \le \bar R < 2k^4+3k$.
 
Taking a subsequence $\{n'\}$ of $\{n\}$ if necessary and 
denoting $\{n'\}$ by $\{n\}$, we may assume that
$\lim_{n\to\infty}{k\ell/n}=\alpha>0$ and   
$\lim_{n\to\infty}{k\ell'/n}=\beta>0$. By the assumption   
$$L:=\lim_{h\to\infty} \frac{\Sigma(x_1x_2\cdots x_h)}{h^3} >0$$
holds for $h=n+k(\ell+2\ell')$, $h=n+k(\ell+\ell')$ and $h=n+k\ell$. 
Dividing \eqref{3.2} by $n^3$ and letting $n\to\infty$, we have 
$$
L (1+\alpha+2\beta)^3- 2L(1+\alpha+\beta)^3
+ L (1+\alpha)^3= \frac{2\alpha\beta^2}{k}+\frac{2\beta^3}{k}.
$$
Since $\ell$, $\ell'$ can be arbitrary satisfying (\ref{1}), (\ref{2}) above, 
this should holds for any $\alpha, \beta >0$ with 
$\alpha+2\beta<A(1-\epsilon)$, which is 
impossible since the lefthand side is
$6 L ((1+\alpha)\beta^2+\beta^3)$ and has a 
term of $\beta^2$ which righthand side hasn't.
\prfend

\noindent{\bf Acknowledgement:}~The authors thank 
Dongguk University for inviting one of the authors and
giving opportunity of the collaboration.

\noindent Teturo Kamae\\
Osaka City University Advanced Mathematics Institute\\
Osaka, 558-8585 Japan\\
(e-mail)~kamae@apost.plala.or.jp\\
(home page)~http://www14.plala.or.jp/kamae\vspace{1em}\\
Dong Han Kim\\
Department of Mathematics Education, Dongguk University - Seoul,\\
Seoul 100-715, Republic of Korea\\
(e-mail)~kim2010@dongguk.edu

\end{document}